# Magnetic states and optical properties of single-layer carbon-doped hexagonal boron nitride

*Hyoungki Park,*[*][†] *Amita Wadehra,*[†] *John W. Wilkins,*[†] *and Antonio H. Castro Neto*[‡]

[†]Department of Physics, The Ohio State University, Columbus, Ohio 43210, USA

[‡]Graphene Research Centre, National University of Singapore, 6 Science Drive 2, Singapore 117546

Abstract: We show that carbon-doped hexagonal boron nitride (h-BN) has extraordinary properties with many possible applications. We demonstrate that the substitution-induced impurity states, associated with carbon atoms, and their interactions dictate the electronic structure and properties of C-doped h-BN. Furthermore, we show that stacking of localized impurity states in small C clusters embedded in h-BN forms a set of discrete energy levels in the wide gap of h-BN. The electronic structures of these C clusters have a plethora of applications in optics, magneto-optics, and opto-electronics.

The advent of the field of two-dimensional (2D) crystals is marked by the isolation and electronic characterization of graphene.[1-3] Not long after that, other layered 2D crystals were also isolated, either by mechanical means,[4, 5] or by artificial synthesis,[6-8] in the search for 2D analogues of 3D materials with technological importance.[9] One such material of interest is hexagonal boron nitride (h-BN), an insulating material that serves as an excellent dielectric substrate for graphene electronics. We study, using state of the art *ab initio* methods, electronic and magnetic properties of single-layer carbon (C) doped h-BN. We show that accurate insights into the interactions mediated by the impurity states pertaining to substitutional C atoms in h-BN can pave the way for engineering of interesting magnetic and optical properties in these 2D materials. Given that these 2D crystals are atomically thin films that are easily accessible by a variety of experimental probes, and the fast development in the controlled synthesis of these materials,[10, 11] a new venue for exploration in quantum computation, photonics, magneto-optics, and opto-electronics based on 2D crystals is unraveled.

The capability of modeling and predicting the electronic properties of crystals is one of the greatest theoretical achievements of the 20[th] century. Given the novelty of 2D crystals and the lack of experimental data, theoretical modeling plays an important role in predicting the feasibility and usefulness of unexplored structures, and hence providing a route to their engineering and manipulation. However, the predictive power of theoretical methods crucially relies on their accuracy. Recent studies[12-18] of C doped h-BN have used density functional theory (DFT) with standard functionals such as Perdew-Burke-Ernzerhof (PBE) functional.[19] However, standard DFT severely underestimates band gaps. Hybrid functionals overcome this problem by including a fraction of the Hartree-Fock exchange energy. One such functional is the Heyd-Scuseria-Ernzerhof (HSE) functional,[20] which we have successfully used to study important properties of technologically important semiconductors.[21-25] Here we present a comprehensive HSE study for electronic structure of C clusters embedded in the single-layer h-BN.

The calculated band gap of the pure single-layer h-BN is a clear demonstration of the accuracy of the HSE functional. HSE computes a gap of 5.69 eV, which is in excellent agreement with experiments,[26, 27] whereas PBE yields the band gap of 4.66 eV underestimating the gap value by approximately 1 eV. Similarly, HSE and PBE functionals produce band gaps of 5.57 eV and 4.22 eV, respectively, for bulk h-BN that has the graphite-like 3D layered structure. These results reassure that HSE functional is more suitable for describing semiconducting 2D crystals.

Given the large gap value in h-BN, it is commonly assumed as a sort of electronic "vegetable" and has been used only as a substrate for graphene. The largest electron mobilities in supported graphene have been obtained with BN as the substrate.[28] Here we show that even small quantities of C atoms can change this picture drastically and transform BN to be an amazing playground for 2D physics.

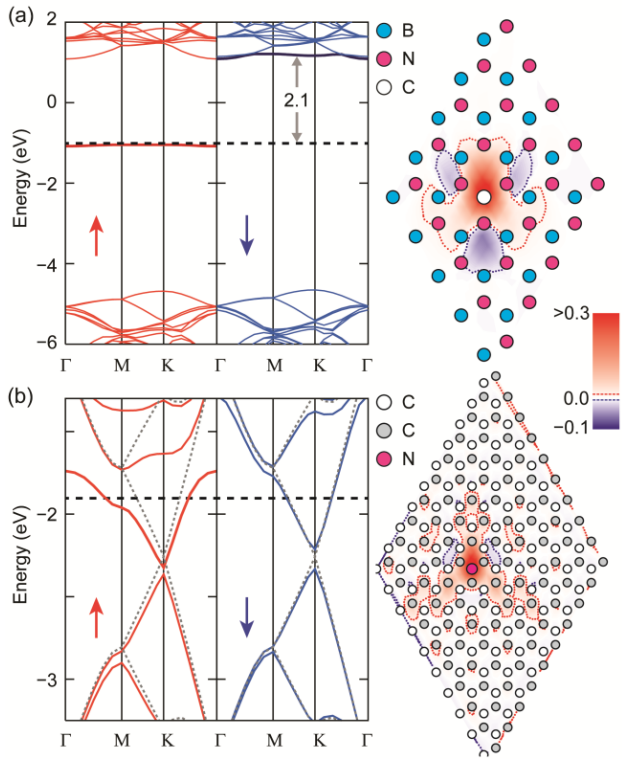

**Figure 1.** Spin-resolved band structure and spin density plots for single substitutions in h-BN and graphene. (a) The impurity state associated with the C atom replacing a B atom is located close to the conduction band of h-BN. A significant spin-splitting results in occupied spin-up (red) and unoccupied spin-down (blue) state, yielding a net magnetic moment of 1 $\mu_B$. The spin density is highly localized around the substitutional atom in h-BN. For the case of a substitutional N in graphene, (b), the partial occupation of both spin-up and spin-down components of the hybridized impurity band results in a very small net magnetic moment. The resultant spin density is highly delocalized and a staggered spin-ordering is observed between two graphene sublattices. For (a), the 50-atom super-cell, 5 × 5 of the two atom unit-cell, is used, but 200-atom 10 × 10 super-cell is used for (b). The black dashed line marks the Fermi level and gray dotted lines represent bulk graphene bands.

For systems with a small concentration of C atoms embedded in 2D h-BN, any changes in the electronic structure are likely to be localized in the vicinity of C atoms due to the large band gap. This is clearly illustrated in our results, Figure 1(a), for an isolated single C atom substituting either a B or a N atom. The substitution gives rise to an unpaired electron, and induces a local magnetic moment. Given that the spin-orbit coupling in BN is weaker than in diamond because of the smaller $sp^3$ distortion, the spin relaxation times are expected to be much longer.[29]

Figure 1(a) shows spin-resolved band structure and spin density for an isolated C atom substituting a B atom in h-BN. The flat impurity state associated with the C atom acts as a donor level. The splitting between spin-up (red) and spin-down (blue) impurity states exceeds 2 eV, which ensures occupation of the spin-up state. When a C replaces a N atom, the impurity state acts as an acceptor level and the occupied spin-up state lies just above the valence band. There is no spin-compensation and we obtain a magnetic moment of 1 $\mu_B$ (S=1/2). As evident from the right hand side plot of Fig. 1(a), the substitutional atom develops a tightly localized spin density. The tail of the spin density barely reaches fourth nearest neighbor atoms (~0.5 nm) indicating that the spin-spin interaction between C atoms is short-ranged. The site-projected wave-function character of the impurity state shows the same distance dependence. In contrast, PBE functional yields a very small spin-splitting of 0.9 eV.

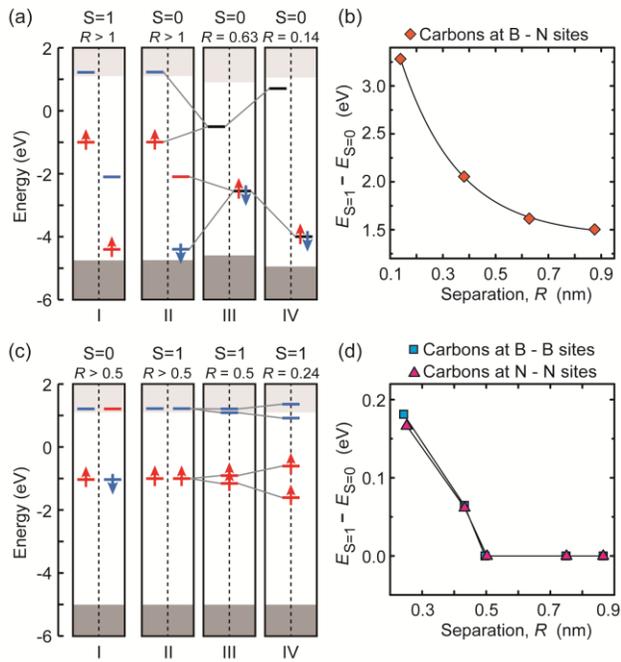

**Figure 2.** Impurity state levels and trends of energy difference between S=0 and S=1 spin-state for 2-D h-BN with carbon substitutions. (a) and (b), Two C atoms are at A-B sublattice sites – substituting a B and a N atom, and hence inducing donor- and acceptor-like flat impurity states deep in the h-BN gap of 5.69 eV. (c) and (d), Both C atoms are substituting B (or N) atoms, the A-A interaction. In both cases, there are four energy spectrum panels, and the exchange trend as a function of separations. Each spectrum panel is divided by a dashed line into two, representing states for each of two C atoms. Blue and red line segments are up and down spin states,

respectively, and black segments are corresponding to spin-degenerate states. If both C atoms contribute equally to the formation of states, the states are drawn across the dashed line and the center of the line segment being at the dashed line.

For comparison and increasing interest in 2D spintronic devices, we investigate the substitution-induced magnetic moments in graphene. In Fig. 1(b), at the $\Gamma$-point, the spin-up impurity state induced by substituting a C atom in graphene by a N atom is located about 0.7 eV below the lowest conduction state of graphene, and shows no sign of hybridization with states from graphene. However, near the K-point, it induces a gap at the "Dirac point" by breaking the sub-lattice symmetry,[30] hybridizes with the delocalized $\pi^*$-orbital, and shows the delocalized character. The unpaired electron occupies the spin-up impurity state and raises the Fermi level of the system by $\sim$ 0.5 eV. On the other hand, the spin-down impurity state lies about 0.5 eV above the spin-up state at the $\Gamma$-point and is closer to the conduction band of graphene. The spin-down state hybridizes with graphene states along all the high-symmetry $k$-points. The hybridized states near the $\Gamma$-point are far above the Fermi level and hence unoccupied, whereas those near the K-point lie below the Fermi energy and show partial occupation. The down-spin electronic structure of graphene near the Fermi level is hardly affected by the substitution with a N atom. The net magnetic moment induced by the unpaired electron is small in graphene, and the spin density for this system is extended, in strong contrast to the highly localized spin density for a substitutional C atom in h-BN. Hence, impurity states in graphene are resonances with a finite lifetime and we expect to observe shorter spin relaxation times than those for the C atoms in BN.

In h-BN, with increased C atom concentration, the local electronic structure is affected by interaction between C atoms. Understanding this local spin-spin interaction is crucial for the description of magnetism in this 2D material. Figure 2 shows spin-polarized electronic structures of two interacting C atoms. Depending on sublattices that they belong to, the two C atoms interact differently. Figure 2(a) presents the case where one C atom is at B site and the other is at N site − the A-B interaction, whereas in Fig. 2(c), two C atoms are at the same sublattice sites − the A-A interaction. The energy difference between the singlet (S=0) and the triplet (S=1) states is presented. When two C atoms are separated beyond their interaction range, the S=0 and the S=1 spin configurations are energetically indistinct. The first two panels (panel I and II) of Fig. 2(a) and (c) show spin-resolved impurity state energy levels for the S=0 and S=1 states with two isolated C atoms at A-B and A-A sublattice sites, respectively. Moving C atoms closer lifts the degeneracy and the S=0 state becomes energetically favorable.

In the case of A-B interactions, when C atoms are brought within the "interaction range", a "tunneling" of the electron from the B-site C atom to the N-site C atom is triggered by the difference in "electronegativity" between two C atoms in different local surroundings. This induces the collapse of spin-splitting of impurity states and results in the S=0 configuration. Impurity states for the N-site C atom with two electrons relaxes to be a spin-degenerate state at about -2.5 eV and the unoccupied state that belongs to the B-site C atom about 2.0 eV above it (Fig. 2(a), panel III). This is analogous to the formation of an "ionic bond". Bringing C atoms even closer increases the overlapping of wavefunctions of these states and leads to hybridization that shifts the two spin-degenerate impurity states away from each other. The energy difference between the occupied and the unoccupied spin-degenerate impurity states is widened to be 4.7 eV for the closest separation (Fig. 2(a), panel IV). The S=1 configuration shows the electronic structure that is the superposition of two isolated C atoms with one spin-up electron sitting on each of the two C atoms. Alternatively, the "molecular orbital" of this spin configuration has the "anti-bonding" character. Since the "anti-bonding" orbitals have a node in between two C atoms, the state-hybridization due to the wavefunction overlapping is negligible. Energy differences between S=1 and S=0 configurations are a few eV and exponentially decrease with separations (Fig. 2(b)). Based on the extent of the spin-density and the wavefunction we estimate that the "interaction range" for this A-B type substitution is about 1 nm.

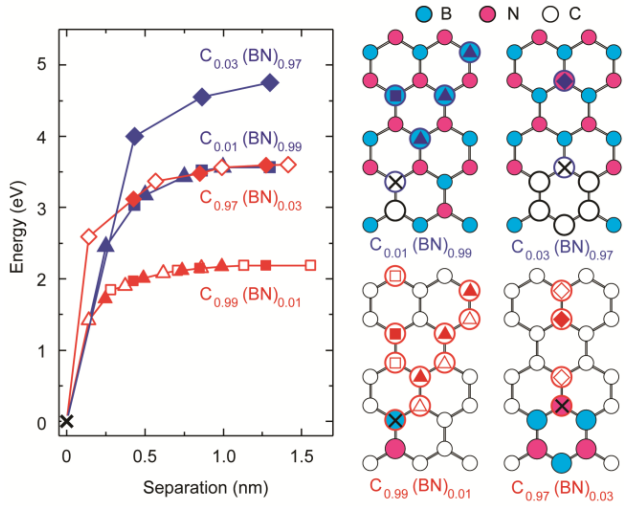

**Figure 3.** Changes in formation energy with increasing separation of a B (C) atom from the position of a B-N pair (C-C pair) or a B-N (C) hexagonal-ring embedded in graphene (h-BN). The energy of a pair or the closed hexagonal ring configuration is set to be zero. The structures shown on the right side are representative of the actual $10 \times 10$ supercell (200 atoms) used for calculation. As schematically displayed, we refer to the position change of the B (C) atom marked by a cross from the pair configuration to one of sites labeled by triangles or squares as the "zigzag" or "armchair" separation, respectively. For hexagonal-rings, only "armchair" separations are studied and sites are labeled by diamonds. Filled and open symbols correspond to two inequivalent-sublattice sites of the 2D hexagonal lattice.

Two C atoms at the same sublattice sites, the A-A case, possess the same local electronic structure when they are far apart. Unlike the A-B case, upon the formation of S=0 state, the A-A interaction does not invoke a charge transfer because both sites have same "electronegativity". The S=0 state consists of the spin-up electron at one C atom and the spin-down electron at the other C atom, and shows the "molecular orbital" with the "anti-bonding" character (Fig. 2(c), panel I). Hence, energy levels of impurity states show no dependence on the separation of C atoms. On the other hand, in the S=1 configurations, two spin-up wavefunctions overlap and split into two hybridized states, which show the "bonding" character. Two C atoms equally share electrons at these hybridized states. The size of the splitting due to the hybridization depends on the separation. The closest separation shows about 1 eV splitting for the S=1 hybridized states (Fig. 2(c) panel IV). Energy differences between the S=1 and the S=0 configurations are an order of magnitude smaller than the A-B case, and the "interaction range" is about 0.5 nm (Fig. 2(d)).

Recent experimental investigations on 2D-BCN materials reveal that graphene and h-BN form domains but do not fully phase-segregate.[11] Figure 3 shows, in terms of formation energy, that both C and BN have a very strong tendency to form pairs and clusters as opposed to being randomly distributed, in agreement with the experimental observations.[10] As seen in Fig. 3, all energies rise rapidly with small separations and level off for large separations. Breaking up a B-N pair embedded in graphene requires at least 1.4 eV, and about 2.2 eV for large separation. Similarly, displacing a N atom from a B-N hexagonal ring costs 2.6–3.6 eV. Even higher energies are needed to separate a C atom from a C pair or from a C hexagonal ring embedded in 2D h-BN, 2.5–3.6 eV or 4.0–4.8 eV, respectively. The "zigzag" and the "armchair" types of separations show no distinction in their energy trends for the C-C pair in h-BN or B-N pair in graphene. Likewise, equivalent and inequivalent sub-lattice site separations follow the same trend for B-N in graphene. Our calculations on graphene reveal that electronic structures for the equivalent and the inequivalent sub-lattice site separations are different mainly near the K-point around the Fermi level. Such a localized small difference in the electronic structure is not reflected in the total energy trend.

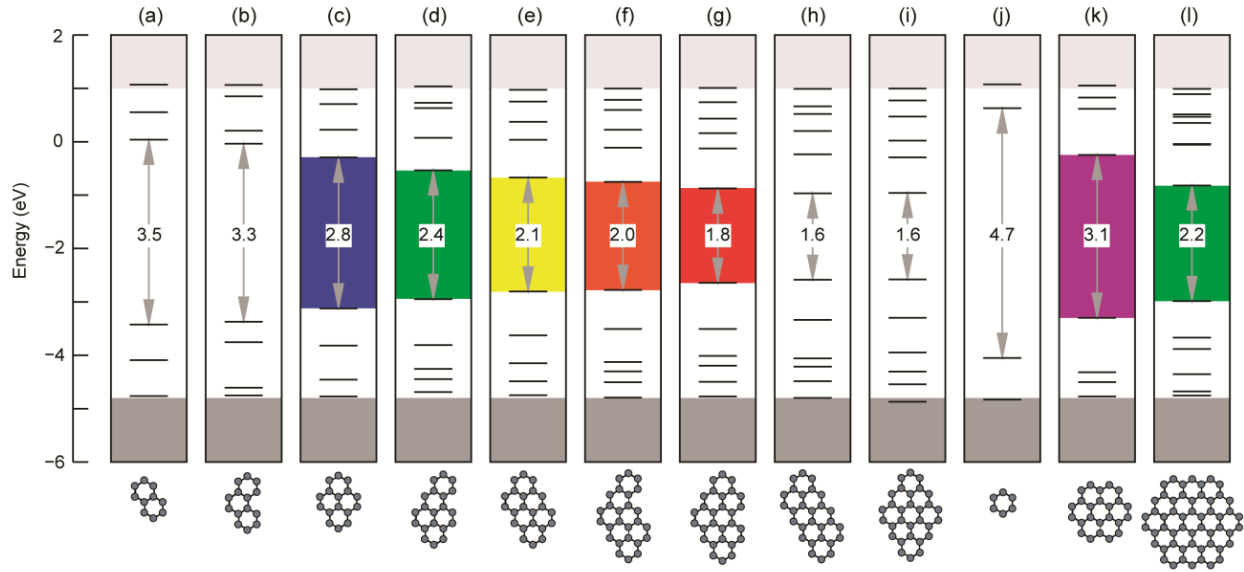

**Figure 4.** Energy spectra for graphene quantum dots embedded in h-BN. (a)-(i) present gradually changing spectra of a 10 × 10 supercell of h-BN containing up to nine C hexagonal rings, whereas (j)-(l) pertain to the six-fold symmetric hexagonal ring configurations containing one, seven, and nineteen rings. The flat energy levels lying between 1.0 eV and -4.8 eV can be ascertained as originating purely from the C quantum dots, and increase with an increasing number of C rings. The levels above 1.0 eV and below -4.8 eV show dispersion due to hybridizations with BN lattice and are contained with the rest of h-BN bands, represented as dark gray (filled) band and light gray (unfilled) band, respectively. The energy difference between the lowest unoccupied level and the highest occupied level, as indicated in the figure, gets reduced with clustering of rings. The energies for (a)-(b) structures lie in the UV range, those for (c)-(g)

gradually span the blue-to-red visible spectrum, whereas (h)-(i) belong to the infrared region. For the six-fold symmetric configurations, multiple doubly-degenerate levels are obtained. The structure containing one C ring, (j), lies within the UV range, whereas the two other symmetric configurations with large number of rings, (k)-(l), are both contained within the visible range of the electromagnetic spectrum.

Figure 4 systematically explores the energy spectra of a variety of C clusters with varying number of hexagonal rings. Our HSE calculations predict that electronic gaps can be gradually spanned, matching the entire visible electromagnetic spectrum, depending on the size and symmetry of the C clusters. Increasing number of C rings stacks more states in the gap of h-BN, and reduces the electronic gap. For systems without the six-fold symmetry, the gaps for smaller clusters containing one to three hexagonal-C rings lie in the UV range, whereas clusters with more than eight rings have gaps belonging to the IR range. On the other hand, the six-fold symmetry allows multiple degenerate states, and gaps for symmetric C clusters vary relatively slowly with increasing number of C rings. For large clusters, we find that the energy gap, $\Delta$, scales perfectly with what is expected by quantum electronic confinement in graphene,[2] that is, $\Delta \cong \hbar v_F/L$, where $v_F$ ($\sim 10^6$ m/s) is the Fermi velocity in bulk graphene and L is the cluster size. Our conclusion is, therefore, that depending on the C concentration and cluster size, the color of h-BN can be modified from the infrared to the ultraviolet, covering the entire visible spectrum. In fact, by observing fluorescence of C-doped h-BN films it should be possible to estimate the C content. This makes the C-doped h-BN an exceptional material for optical applications. Theoretical studies[12-18] using DFT-PBE have investigated the properties of limited and highly symmetric quantum dots in h-BN. However, PBE values of energy gaps and spin-splitting are lower than HSE results by 0.5–1.0 eV. Hence, even at the qualitative level, the substantial difference between HSE and PBE can lead to completely different conclusions on many material properties.

We use the plane-wave projector augmented-wave (PAW) method[31] with the HSE06 hybrid functional in the vasp code.[32-34] We use a plane-wave energy cutoff of 700 eV for all our calculations. The hybrid BCN nanostructures are modeled in 200 atom super-cells, $10 \times 10$ of the two atom unit-cell. The Brillouin zone integration for graphene and the single-layer h-BN is performed on a $\Gamma$-centered $10 \times 10 \times 1$ k-point meshes for the unit-cell, and equivalent meshes for larger simulation cells. Band structures are computed on discrete k-point meshes along high symmetry directions: $\Gamma$-M-K-$\Gamma$.

In summary, using state of the art ab initio methods we have shown that C-doped h-BN shows extraordinary properties with possible applications in many different fields of condensed matter research. We demonstrate that the substitution-induced impurity states, associated with carbon atoms, and their interactions dictate the electronic structure and properties of C-doped h-BN. Hence, an accurate description of these impurity states is essential to understand and engineer properties of this new class of 2D materials. Stacking of localized impurity states in small C clusters embedded in h-BN forms a set of discrete energy levels in the wide gap of h-BN. The electronic structures of these C clusters have a plethora of applications in optics, magneto-optics, and opto-electronics. Hence, our results open a new venue for the study of complex magnetic and optical phenomena in 2D crystals.

AUTHOR INFORMATION

**Corresponding Author**

*Email: hkpark@mps.ohio-state.edu.

**Notes**

The authors declare no competing financial interest.

ACKNOWLEDGMENT

This work was supported by DOE-BES-DMS (DEFG02-99ER45795). We used computational resources of the NERSC, supported by the U.S. DOE (DE-AC02-05CH11231), and the OSC. AHCN acknowledges DOE grant DE-FG02-08ER46512, ONR grant MURI N00014-09-1-1063, and the NRF-CRP award "Novel 2D materials with tailored properties: beyond graphene" (R-144-000-295-281).

REFERENCES

1.    Novoselov, K. S.; Geim, A. K.; Morozov, S. V.; Jiang, D.; Zhang, Y.; Dubonos, S. V.; Grigorieva, I. V.; Firsov, A. A. *Science* **2004,** *306*, 666-669.

2.    Castro Neto, A. H.; Guinea, F.; Peres, N. M. R.; Novoselov, K. S.; Geim, A. K. *Rev. Mod. Phys.* **2009,** *81*, 109-162.

3.    Geim, A. K.; Novoselov, K. S. *Nat. Mater.* **2007,** *6*, 183-191.

4.      Novoselov, K. S.; Jiang, D.; Schedin, F.; Booth, T. J.; Khotkevich, V. V.; Morozov, S. V.; Geim, A. K. *PNAS* **2005,** *102*, 10451-10453.

5.      Radisavljevic, B.; Radenovic, A.; Brivio, J.; Giacometti, V.; Kis, A. *Nature Nanotech.* **2011,** *6*, 147-150.

6.      Elias, D. C.; Nair, R. R.; Mohiuddin, T. M. G.; Morozov, S. V.; Blake, P.; Halsall, M. P.; Ferrari, A. C.; Boukhvalov, D. W.; Katsnelson, M. I.; Geim, A. K.; Novoselov, K. S. *Science* **2009,** *323*, 610-613.

7.      Nair, R. R.; Ren, W.; Jalil, R.; Riaz, I.; Kravets, V. G.; Britnell, L.; Blake, P.; Schedin, F.; Mayorov, A. S.; Yuan, S.; Katsnelson, M. I.; Cheng, H.-M.; Strupinski, W.; Bulusheva, L. G.; Okotrub, A. V.; Grigorieva, I. V.; Grigorenko, A. N.; Novoselov, K. S.; Geim, A. K. *Small* **2010,** *6*, 2877-2884.

8.      Kim, K. S.; Zhao, Y.; Jang, H.; Lee, S. Y.; Kim, J. M.; Kim, K. S.; Ahn, J.-H.; Kim, P.; Choi, J.-Y.; Hong, B. H. *Nature* **2009,** *457*, 706-710.

9.      Castro Neto, A. H.; Novoselov, K. S. *Rep. Prog. Phys.* **2011,** *74*, 082501.

10.     Ci, L.; Song, L.; Jin, C. H.; Jariwala, D.; Wu, D. X.; Li, Y. J.; Srivastava, A.; Wang, Z. F.; Storr, K.; Balicas, L.; Liu, F.; Ajayan, P. M. *Nat. Mater.* **2010,** *9*, 430-435.

11.     Shi, Y.; Hamsen, C.; Jia, X.; Kim, K. K.; Reina, A.; Hofmann, M.; Hsu, A. L.; Zhang, K.; Li, H.; Juang, Z.-Y.; Dresselhaus, M. S.; Li, L.-J.; Kong, J. *Nano Lett.* **2010,** *10*, 4134-4139.

12.     Lam, K.-T.; Lu, Y.; Feng, Y. P.; Liang, G. *Appl. Phys. Lett.* **2011,** *98*, 022101.

13.     Li, J.; Shenoy, V. B. *Appl. Phys. Lett.* **2011,** *98*, 013105.

14.   Ramasubramaniam, A.; Naveh, D. *Phys. Rev. B* **2011,** *84*, 075405.

15.   Kan, M.; Zhou, J.; Wang, Q.; Sun, Q.; Jena, P. *Phys. Rev. B* **2011,** *84*, 205412.

16.   Ramasubramaniam, A.; Naveh, D.; Towe, E. *Nano Lett.* **2011,** *11*, 1070-1075.

17.   Manna, A. K.; Pati, S. K. *J. Phys. Chem. C* **2011,** *115*, 10842-10850.

18.   Bhowmick, S.; Singh, A. K.; Yakobson, B. I. *J. Phys. Chem. C* **2011,** *115*, 9889-9893.

19.   Perdew, J. P.; Burke, K.; Ernzerhof, M. *Phys. Rev. Lett.* **1996,** *77*, 3865-3868.

20.   Heyd, J.; Scuseria, G. E.; Ernzerhof, M. *J. Chem. Phys.* **2003,** *118*, 8207-8215.

21.   Hennig, R. G.; Wadehra, A.; Driver, K. P.; Parker, W. D.; Umrigar, C. J.; Wilkins, J. W. *Phys. Rev. B* **2010,** *82*, 014101.

22.   Wadehra, A.; Nicklas, J. W.; Wilkins, J. W. *Appl. Phys. Lett.* **2010,** *97*, 092119.

23.   Nicklas, J. W.; Wilkins, J. W. *Appl. Phys. Lett.* **2010,** *97*, 091902.

24.   Nicklas, J. W.; Wilkins, J. W. *Phys. Rev. B* **2011,** *84*, 121308(R).

25.   Park, H.; Wilkins, J. W. *Appl. Phys. Lett.* **2011,** *98*, 171915.

26.   Watanabe, K.; Taniguchi, T.; Kanda, H. *Nat. Mater.* **2004,** *3*, 404-409.

27.   Song, L.; Ci, L.; Lu, H.; Sorokin, P. B.; Jin, C.; Ni, J.; Kvashnin, A. G.; Kvashnin, D. G.; Lou, J.; Yakobson, B. I.; Ajayan, P. M. *Nano Lett.* **2010,** *10*, 3209-3215.

28.   Dean, C. R.; Young, A. F.; Meric, I.; Lee, C.; Wang, L.; Sorgenfrei, S.; Watanabe, K.; Taniguchi, T.; Kim, P.; Shepard, K. L.; Hone, J. *Nature Nanotech.* **2010,** *5*, 722-726.

29.   Castro Neto, A. H.; Guinea, F. *Phys. Rev. Lett.* **2009,** *103*, 026804.

30.   Pereira, V. M.; dos Santos, J. M. B. L.; Castro Neto, A. H. *Phys. Rev. B* **2008,** *77*, 115109.

31.   Blochl, P. E. *Phys. Rev. B* **1994,** *50*, 17953-17979.

32.   Kresse, G.; Furthmuller, J. *Phys. Rev. B* **1996,** *54*, 11169-11186.

33.   Kresse, G.; Joubert, H. *Phys. Rev. B* **1999,** *59*, 1758-1775.

34.   Paier, J.; Marsman, M.; Hummer, K.; Kresse, G.; Gerber, I. C.; Angyan, J. G. *J. Chem. Phys.* **2006,** *124*, 154709.

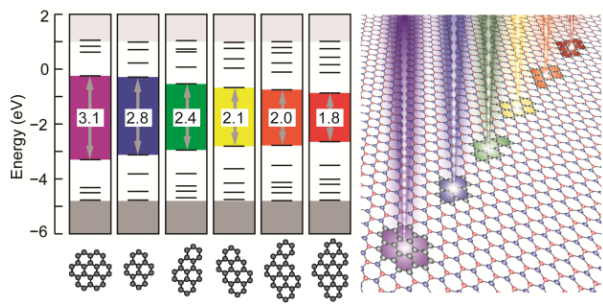